\documentclass[preprint]{imsart}
\usepackage{amsmath,amssymb,amsfonts,graphicx,amsthm,nicefrac,mathtools,bm}
\usepackage[numbers]{natbib}
\usepackage[english]{babel}
\usepackage{times}
\usepackage[T1]{fontenc}
\usepackage{bbm}
\usepackage{color}
\usepackage{xspace}
\usepackage{rotating,multirow}

\startlocaldefs

\theoremstyle{definition}

\theoremstyle{remark}

\makeatletter
\newtheorem*{rep@theorem}{\rep@title}
\newcommand{\newreptheorem}[2]{%
\newenvironment{rep#1}[1]{%
 \def\rep@title{#2 \ref{##1}}%
 \begin{rep@theorem}}%
 {\end{rep@theorem}}}
\makeatother
\newreptheorem{theorem}{Theorem}
\newreptheorem{lemma}{Lemma}
\newreptheorem{corollary}{Corollary}
\newreptheorem{proposition}{Proposition}

\newcommand{\ignore}[1]{}

\newcommand{\bea}{\begin{eqnarray}}
\newcommand{\eea}{\end{eqnarray}}
\newcommand{\beaa}{\begin{eqnarray*}}
\newcommand{\eeaa}{\end{eqnarray*}}

\renewcommand{\bar}{\overline}

\renewcommand{\hat}{\widehat}
\renewcommand{\tilde}{\widetilde}

\renewcommand{\mid}{\mbox{\,$|$\,}}

\newcommand{\bmZ}{\boldsymbol{Z}}

\newcommand{\bmzero}{\boldsymbol{0}}
\newcommand{\bmbeta}{\boldsymbol{\beta}}

\endlocaldefs

\begin{document}

\begin{frontmatter}

\title{On a Shape-Invariant Hazard Regression Model}
\runtitle{On a Shape-Invariant Hazard Regression Model}


\author{\fnms{Cheng} \snm{Zheng}\ead[label=e1]{zhengc@uwm.edu}}
\address{Joseph J. Zilber School of Public Health, University of Wisconsin-Milwaukee, Milwaukee WI \\ \printead{e1}}
\and
\author{\fnms{Ying Qing} \snm{Chen}\ead[label=e2]{yqchen@fhcrc.org}}
\address{Program in Biostatistics and Biomathematics, Fred Hutchinson Cancer Research Center, Seattle WA \\ \printead{e2}}
\runauthor{C. Zheng and Y. Q. Chen}

\begin{abstract}
In survival analysis, Cox model is widely used for most clinical trial data. Alternatives include the additive hazard model, the accelerated failure time (AFT) model and a more general transformation model. All these models assume that the effects for all covariates are on the same scale. However, it is possible that for different covariates, the effects are on different scales. In this paper, we propose a shape-invariant hazard regression model that allows us to estimate the multiplicative treatment effect with adjustment of covariates that have non-multiplicative effects. We propose moment-based inference procedures for the regression parameters. We also discuss the risk prediction and goodness of fit test for our proposed model. Numerical studies show good finite sample performance of our proposed estimator. We applied our method to Veteran's Administration (VA) lung cancer data and the HIVNET 012 data. For the latter, we found that single-dose nevirapine treatment has a significant improvement for 18-month survival with appropriate adjustment for maternal CD4 counts and virus load. 

\end{abstract}



\end{frontmatter}

\section{Introduction}
\label{intro}

An important HIV/AIDS randomized prevention trial was conducted between November 1997 and January 2001 (Jackson \emph{et al.} 2003). This trial was named as HIVNET 012 and the goal was to evaluate the efficacy and the safety of a short course nevirapine (NVP) treatment compared to a short course zidovudine (AZT) treatment for pregnant mothers during labor and delivery. The primary clinical endpoint is the mother-to-child transmission (MTCT) of human immunodeficiency virus type-1 (HIV-1) and the secondary endpoint is the 18-months infant survival. As is shown in Chen \textit{et al.} (2012), the NVP seems improving survival over time when looking at the Kaplan-Meier curve. Cox regression analysis also suggests that the NVP would reduce hazard by 26.0\% but this improvement is not statistically significant (p$>$0.20). So it is natural to ask whether there is alternative models to detect the treatment effect. For this data, we considered two important covariates, maternal HIV-RNA viral loads (VLs) and maternal CD4+ counts at baseline. However, we did not restrict the covariate to be time-independent and follow-up VLs and CD4+ counts can be used when trying to identify the mechanism how the treatment works. Another possible time-dependent covariate, for example, is the interaction between time and treatment assignment.

In survival analysis of censored time-to-event outcomes, Cox model (Cox 1972) is most commonly used due to its simple mathematical structure in terms of its parameter estimation, asymptotic result etc. The additive hazard model (Lin and Ying 1994) was proposed so as the covariates' effect on hazard is treated as additive, rather than multiplicative. An alternative model, the accelerated failure time model (AFT; Zeng and Lin 2007), is commonly used in the field of engineering or physics (i.e. reliability assessment) given the better scientific explanation of its coefficients. However, all the models above require that different covariates have the same type of effects on the unified transformed scale. It is possible that different risk factors have different kinds of effects; modeling all variables' effects in the same scale might be inappropriate.

We denote the treatment indicator as $Z$ and denote the two covariates as $X_1$ and $X_2$. The traditional Cox regression model has the following hazard function form:
\begin{equation}
\lambda(t\mid Z,X_1,X_2)=\lambda_0(t)\exp\{\gamma Z+\beta_1X_1+\beta_2X_2\},
\end{equation}
where $\lambda_0(\cdot)$ is an unknown baseline hazard function and $\gamma$, $\beta_1$, $\beta_2$ are parameters of interest. One important feature of this model is that the marginal hazard, as well as the marginal survival curve, cannot cross-over between different groups defined by covariates. However, for real data, the Kaplan-Meier curves often have some cross-overs and this indicates that these simple models may not be sufficient.

Several models have been proposed to model different kinds of effects. Lin and Ying (1995) proposed an additive-multiplicative hazard model that allows two kinds of effects. Chen and Jewell (2001) proposed an accelerated hazard model which is a flexible model allowing effects in both multiplicative hazard scale and time scale change. Combining the effects allowed in these two models, we have the following hazard form:
\begin{equation}
\lambda(t\mid Z,X_1,X_2)=\lambda_0(t e^{\beta_1 X_1})e^{\beta_1X_1}e^{\gamma Z}+\beta_2X_2.
\end{equation}
We use this model since it allows the survival functions for different levels of $Z$ to have cross-over, which is the case for many real data problem.

In this paper, we propose a broader class of model that allows all three kinds of effects. To make the interpretation of each parameter clear, our method focus on the cases where covariates are different for different effects in the main part of the paper followed by a discussion on the extension to allow same covariate have different kinds of effects. Besides estimating the effect of certain covariate, our model is useful to make prediction because it is closer to a nonparametric model and has fewer restrictions for the survival curves between different groups defined by covariates.

We organize the paper as follows. In section 2, we propose a shape invariant hazard model. In section 3, we present the inference method with asymptotic results. In section 4, we perform numerical study to evaluate the finite sample performance of our estimator. In section 5, we apply our method to the VA lung cancer data and HIVNET 012 data. In section 6, we provide more discussion of the model.

\section{The Shape-Invariant Hazard Regression Model}
\label{sec:1}
To make model (2) general, we, mathematically, treat $Z$, $X_1$ and $X_2$ all as covariate. We denote $\bmZ=(\bmZ_1,\bmZ_2,\bmZ_3)$, and propose the following ``Shape-Invariant" hazard model:
\begin{equation}
\lambda(t\mid \bmZ)=\lambda_0(t e^{\bmbeta_1 \bmZ_1})e^{\bmbeta_1\bmZ_1}e^{\bmbeta_2\bmZ_2}+\bmbeta_3\bmZ_3,
\end{equation}
where $\bmbeta=(\bmbeta_1,\bmbeta_2,\bmbeta_3)$ are unknown parameters of interest.

When $\bmZ_2$ and $\bmZ_3$ are potentially time-varying, we propose the following model:
\begin{equation}\label{model.sim}
\lambda(t\mid \bmZ)=\lambda_0(t e^{\bmbeta_1 \bmZ_1})e^{\bmbeta_1\bmZ_1}e^{\bmbeta_2\bmZ_2(t)}+\bmbeta_3\bmZ_3(t).
\end{equation}

Here $\bmZ_1$, $\bmZ_2(t)$, $\bmZ_3(t)$ are assumed to be different. So for this model, we have an accelerated time effect from $\bmZ_1$, then a multiplicative hazard effect from $\bmZ_2(t)$ and finally an additive effect from $\bmZ_3(t)$. An important feature for this model is that for any subgroup defined by $\bmZ_1$ and $\bmZ_2$, the model is an additive risk model with respect to $\bmZ_3$. But the model is not a Cox model for some subgroups defined by $\bmZ_1$ and $\bmZ_3$. It is only a Cox model respect to $\bmZ_2$ for the subgroup with $\bmZ_3=0$. Also, the model is an AFT model respect to $\bmZ_1$ only in the subgroup with $\bmZ_2=0$ and $\bmZ_3=0$. 

We denote the derivative of the baseline hazard as $\dot{\lambda}_0(t)$. To make $\bmbeta$ in model (4) identifiable, we need that for some $t$,
$$
[\lambda_0(t e^{\bmbeta_1 \bmZ_1})+te^{\bmbeta_1 \bmZ_1}\dot{\lambda}_0(t e^{\bmbeta_1 \bmZ_1})]\bmZ_1e^{\bmbeta_1\bmZ_1+\bmbeta_2\bmZ_2(t)}
$$
$\lambda_0(t e^{\bmbeta_1 \bmZ_1})\bmZ_2(t)e^{\bmbeta_1\bmZ_1+\bmbeta_2\bmZ_2(t)}$, and $\bmZ_3(t)$ are linear independent at true parameter. When the joint distribution of $\bmZ_1, \bmZ_2, \bmZ_3$ are non-degenerate, this assumption holds.

The cumulative hazard function of our proposed model has the following form:
\begin{equation}
\Lambda(t\mid \bmZ)=\int_0^t[e^{\bmbeta_2\bmZ_2(u)}d\Lambda_0(u e^{\bmbeta_1 \bmZ_1})+\bmbeta_3\bmZ_3(u)du],
\end{equation}
where $\Lambda_0(\cdot)$ is the baseline cumulative hazard function. When $\bmZ_2$ is time-independent, which is true when $\bmZ_2$ is treatment indicator, we have
\begin{equation}
\Lambda(t\mid \bmZ)=e^{\bmbeta_2\bmZ_2}\Lambda_0(t e^{\bmbeta_1 \bmZ_1})+\bmbeta_3\int_0^t\bmZ_3(u)du.
\end{equation}

\section{Inference}
\label{sec:2}
In this section, we give moment-based inference procedures for our proposed shape-invariant hazard model. We give the estimator, followed by the asymptotic result. Then, we discuss the prediction and test for goodness fit of the model.
\subsection{Estimation of Survival Parameters}
Here, we estimate the survival parameters with moment-based estimation procedure. Denote the event time as $T$ and the censoring time as $C$, then the observed composite endpoints are $T^{\ast}=T\wedge C$ and $\Delta=I(T\leq C)$. Using traditional counting process notation, we denote $Y_i(t)=I(T^{\ast}_i\geq t)$ and $N_i(t)=I(\Delta_i=1,T^{\ast}_i\leq t)$. We denote our data as $\underline{X}=\{Y_i(t), N_i(t), \bmZ_i(t), t\in [0,\infty), i=1,2,\cdots, n\}$. We define the filtration as
\begin{eqnarray*}
\mathcal{F}_t&=&\sigma\{Y_i(s\exp(-\bmbeta_{10}\bmZ_{1i})), N_i(s\exp(-\bmbeta_{10}\bmZ_{1i})), \bmZ_{1i},\bmZ_{2i}(s),\bmZ_{3i}(s),\\
&&s\in[0,t],i=1,\cdots,n\}.
\end{eqnarray*}
We denote the true parameter as $\bmbeta_{0}=(\bmbeta_{10},\bmbeta_{20},\bmbeta_{30})$. Noticing
\bea
&&E[dN_i(te^{-\bmbeta_{10} \bmZ_{1i}})|\mathcal{F}_{t-};\bmbeta_0]\nonumber \\
&=&Y_i(te^{-\bmbeta_{10} \bmZ_{1i}})[\lambda_0(t)e^{\bmbeta_{20}\bmZ_{2i}(t)}+\bmbeta_{30}\bmZ_{3i}(t)e^{-\bmbeta_{10}\bmZ_{1i}}]dt,
\eea
we can obtain a moment-based estimator for baseline hazard function given regression parameters as
\bea
\hat{\Lambda}_0(t;\bmbeta)=\int_0^t\frac{\sum_i[dN_i(se^{-\bmbeta_1 \bmZ_{1i}})-Y_i(se^{-\bmbeta_1 \bmZ_{1i}})\bmbeta_3\bmZ_{3i}(s)e^{-\bmbeta_1\bmZ_{1i}}ds]}{\sum_iY_i(se^{-\bmbeta_1 \bmZ_{1i}})e^{\bmbeta_2\bmZ_{2i}(s)}}.
\eea
We choose different $W_i(t)$ and denote $\bar{W}(t;\bmbeta)=\frac{\sum_jW_j(t)Y_j(te^{-\bmbeta_1 \bmZ_{1j}})e^{\bmbeta_2\bmZ_{2j}(t)}}{\sum_jY_j(te^{-\bmbeta_1 \bmZ_{1j}})e^{\bmbeta_2\bmZ_{2j}(t)}}$, then we can solve the following unbiased estimating equation,
\begin{equation}
0=\sum_i\int\{[W_i(t)-\bar{W}(t;\bmbeta)][dN_i(te^{-\bmbeta_1 \bmZ_{1i}})-Y_i(te^{-\bmbeta_1 \bmZ_{1i}})\bmbeta_3\bmZ_{3i}(t)e^{-\bmbeta_1\bmZ_{1i}}dt]\},
\end{equation}
to obtain a consistent estimator of $\bmbeta$. When $\bmZ_1$, $\bmZ_2$ and $\bmZ_3$ are different, it is natural to use them as the weight function. For simplicity, we assume the weight does not depend on unknown parameters or empirical distribution of the survival time for whole population.

\textbf{Theorem 1:} Under regularity conditions listed in supplementary materials, we have
\bea
\sqrt{n}(\hat{\bmbeta}-\bmbeta_0)\xrightarrow{\mathcal{D}} N(\bmzero,\Sigma).
\eea

The proof of Theorem 1 are given in the supplementary materials.

\subsection{Model-Based Prediction}
For prediction purposes, similar to the Cox model (Lin et al. 1994), the AFT model (Park and Wei 2003) and the additive risk model (Shen and Cheng 1999), pointwise confidence interval as well as the simultaneous confidence bands can be constructed. For pointwise confidence interval, as we have the central limit theory for $\hat{\beta}$ and $\hat{\Lambda}(t)$, the only thing we need to know is the joint distribution of these two quantities to obtain a standard error estimation $v(t)=\widehat{SE}$ of our estimated cumulative hazard function
\bea
\hat{\Lambda}(t|\bmZ)=\int_0^t [e^{\hat{\bmbeta}_2\bmZ_2(u)}d\hat{\Lambda}_0(ue^{\hat{\bmbeta}_1\bmZ_1})+\hat{\bmbeta}_3\bmZ_3(u)du],
\eea
where the baseline hazard is estimated by $\hat{\Lambda}_0(t)=\hat{\Lambda}_0(t,\hat{\bmbeta})$. When $\bmZ$ is time independent, this can be simplified by
\bea
\hat{\Lambda}(t|\bmZ)=\hat{\Lambda}_0(te^{\hat{\bmbeta}_1\bmZ_1})e^{\hat{\bmbeta}_2\bmZ_2}+\hat{\bmbeta}_3\bmZ_3t.
\eea
Then the point-wise confidence interval can be calculated by $\hat{\Lambda}(t|\bmZ)\pm z_{1-\alpha/2}v(t)$,
where $z$ is the critical value from standard normal distribution and $\alpha$ is significant level. We can obtain $\hat{v}(t)$ from simple Bootstrap method. Similarly, we can have the following type of confidence bands $\hat{\Lambda}(t|\bmZ)\pm c(\alpha)v(t)$,where $c(\alpha)$ is obtained from Bootstrap sample.

\subsection{Model Adequacy Assessment}
We consider Kolmogorov-Smirnov Test and Gill-Schumacher Test to assess our model fit. We can compare the model based fitted cumulative hazard to the nonparametric fitted one. This method is not preferred when $\bmZ$ is time-varying. When $\bmZ$ is time independent, we define the test statistics as
\bea
D(t,\hat{\bmbeta},\bmZ)&=&n^{-1/2}\int_0^tQ(u,\hat{\bmbeta},\bmZ)d[\hat{\Lambda}(u|\bmZ)-\hat{\Lambda}_{NP}(u|\bmZ)]
\eea
where $Q(\cdot,\cdot,\cdot)$ is an arbitrary weight function. Here $\hat{\Lambda}_{NP}(u|\bmZ)$ denotes the cumulative hazard function estimated nonparametrically. One such estimator is the Kernel-Smoothed Kaplan-Meier estimator (Akritas, 1994) in the following form:
\bea
\hat{\Lambda}_{NP}(u|\bmZ)=\int_0^u \frac{n^{-1}\sum_{i=1}^nK_h(\bmZ_i-z)dN_i(t)}{n^{-1}\sum_{i=1}^nK_h(\bmZ_i-z)I(Y_i\geq t)dt},
\eea
where $K_h(x)=K(x/h)/h$. Here $K(\cdot)$ is a given symmetric smooth kernel density function and $h$ is the bandwidth such that $nh^2\rightarrow \infty$ and $nh^4\rightarrow 0$ as $n\rightarrow \infty$. We can use $D_1(\hat{\bmbeta})=sup_{t,\bmZ}|D(t,\hat{\bmbeta},\bmZ)|$ or $D_2(\hat{\bmbeta})=sup_{t}|\int D(t,\hat{\bmbeta},\bmZ)d\hat{F}_{\bmZ}|$ where $\hat{F}_{\bmZ}$ is the empirical distribution of $\bmZ$. A very large test statistics suggests that the model is not adequate to fit the data.

Another test we considered here is the Gill-Schumacher Test. For any weight $W_{1i}(t)$ and $W_{2i}(t)$, we can obtain estimators by solving the score equation $S_1(\bmbeta)=0$ and $S_2(\bmbeta)=0$ to obtain $\hat{\bmbeta}^{(1)}$ and $\hat{\bmbeta}^{(2)}$. It is straightforward to use the test statistics $\hat{\bmbeta}^{(1)}-\hat{\bmbeta}^{(2)}$, but the variance of this statistics is difficult to compute. So using the similar estimator as in Chen (2001), we also consider the score test. Since we have that under the null, $n^{-1/2}(S_1(\bmbeta_0),S_2(\bmbeta_0))$ follows joint normal distribution with covariance matrix $\Sigma_{12}$, which can be estimated using Bootstrap method. So we can use test statistics
\bea
T_{GS}=min_{\beta\in U(\hat{\bmbeta}^{(1)})}\left[\left(\begin{array}{c}S_1(\bmbeta)\\S_2(\bmbeta)\end{array}\right)^T\hat{\Sigma}_{12}^{-1}\left(\begin{array}{c}S_1(\bmbeta)\\S_2(\bmbeta)\end{array}\right)\right],
\eea
which follows $\chi^2_p$ distribution asymptotically to assess model fitting.

\section{Numerical Studies}
\label{sec:3}
\subsection{Numerical Estimation Procedure}
Since the estimating equation contains an integration including $Y(t)$, which does not have a close form, we choose 1000 equal space points to evaluate the integration numerically. Since the estimating equation is not differentiable, we use recursive bisector search for the solution that minimize the absolute value of the estimating equations. We need to estimate the variance based on the linear approximation. However, such approximation contains $\lambda(t)$ and $\dot{\lambda}(t)$, which can only be estimated via kernel smoothing and is not stable with finite sample size. So we use a numerical method as given in Chen and Jewell (2001) to estimate variance. If we denote the normalized estimating equation by $0=S(\bmbeta)$, we estimate the variance matrix $\hat{\Sigma}$ of $S(\bmbeta)$ by $n^{-1}\sum_{i=1}^nS_i(\hat{\bmbeta})^{\otimes 2}$. We consider the following decomposition $\hat{\Sigma}=BB^T$, where $B=(b_1,b_2,\cdots,b_p)$. Then we use the following expression for variance estimator
$$
(S^{-1}(n^{-1/2}b_1)-\hat{\bmbeta},\cdots,S^{-1}(n^{-1/2}b_p)-\hat{\bmbeta})^{\otimes 2}.
$$
\subsection{Simulations}
In order to evaluate the finite sample performance of our proposed estimator, we performed simulations as shown below. We chose the baseline hazard as $1/(1+t)$. We considered $\bmZ=(Z_1,Z_2,Z_3)$ where $Z_1$, $Z_2$ and $Z_3$ independently follow bernoulli distribution with parameter 0.5 or follow standard uniform distributions. The censoring distribution is an exponential distribution with appropriate rate to making about $30\%$ or $50\%$ censoring. We ran 100 simulations for each setting and summarized the bias, empirical SE, mean estimated SE. The simulation results are summarized in Table 1-2. From the tables, we can see that our proposed estimator is unbiased. The estimated SE is generally comparable to the empirical one and the empirical SE increases when the censoring rate increases. 

\begin{table}[!h]
\begin{center}
\begin{tabular}{|cc|ccc|ccc|ccc|}
\hline
\multirow{3}{*}{Censoring} &
\multirow{3}{*}{True $\bmbeta$} &
\multicolumn{9}{c|}{n=200}\\
\cline{3-11}
& &
\multicolumn{3}{c|}{Bias} &
\multicolumn{3}{c|}{Est SE} &
\multicolumn{3}{c|}{Emp SE}\\
\cline{3-11}
&  &$\beta_1$  &$\beta_2$ &$\beta_3$ &$\beta_1$  &$\beta_2$ &$\beta_3$&$\beta_1$  &$\beta_2$ &$\beta_3$\\
\hline
30\%&0,0,0.1	&0.03	&-0.02	&0.01	&0.26	&0.07	&0.03	&0.34	&0.20	 &0.09\\
	&0,0,0.2	&0.04	&-0.02	&0.00	&0.29	&0.08	&0.05	&0.35	&0.21	 &0.11\\
	&0,-0.5,0.1	&0.03	&-0.03	&0.00	&0.29	&0.08	&0.03	&0.37	&0.21	 &0.07\\
	&0,-0.5,0.2	&0.02	&-0.03	&0.00	&0.30	&0.08	&0.03	&0.38	&0.22	 &0.08\\
	&0.5,0,0.1	&0.01	&-0.02	&0.01	&0.27	&0.06	&0.04	&0.31	&0.19	 &0.10\\
	&0.5,0,0.2	&0.01	&-0.02	&0.00	&0.25	&0.07	&0.04	&0.33	&0.20	 &0.13\\
	&0.5,-0.5,0.1	&-0.01	&-0.03	&0.00	&0.29	&0.09	&0.04	&0.35	 &0.20	&0.08\\
	&0.5,-0.5,0.2	&-0.01	&-0.03	&0.00	&0.28	&0.09	&0.05	&0.36	 &0.21	&0.10\\
\hline
50\%&0,0,0.1	&0.05	&-0.03	&0.02	&0.30	&0.09	&0.04	&0.36	&0.24	 &0.12\\
	&0,0,0.2	&0.05	&-0.03	&0.01	&0.33	&0.10	&0.06	&0.37	&0.25	 &0.15\\
	&0,-0.5,0.1	&0.04	&-0.03	&0.01	&0.27	&0.10	&0.03	&0.40	&0.24	 &0.09\\
	&0,-0.5,0.2	&0.04	&-0.03	&0.00	&0.29	&0.09	&0.04	&0.40	&0.26	 &0.12\\
	&0.5,0,0.1	&0.04	&-0.02	&0.04	&0.29	&0.10	&0.05	&0.34	&0.23	 &0.13\\
	&0.5,0,0.2	&0.03	&-0.02	&0.02	&0.26	&0.08	&0.05	&0.34	&0.25	 &0.17\\
	&0.5,-0.5,0.1	&0.01	&-0.03	&0.02	&0.23	&0.07	&0.03	&0.35	 &0.23	&0.11\\
	&0.5,-0.5,0.2	&0.02	&-0.03	&0.01	&0.31	&0.09	&0.05	&0.35	 &0.25	&0.14\\
\hline
\end{tabular}
\end{center}
\tabcolsep 5mm \caption{Simulation results for sample size 200 for the baseline hazard $1/(1+t)$, binary covariates, with different true $\bmbeta$ and censoring rate from 100 simulations}
\end{table}
\begin{table}[!h]
\begin{center}
\begin{tabular}{|cc|ccc|ccc|ccc|}
\hline
\multirow{3}{*}{Censoring} &
\multirow{3}{*}{True $\bmbeta$} &
\multicolumn{9}{c|}{n=500}\\
\cline{3-11}
& &
\multicolumn{3}{c|}{Bias} &
\multicolumn{3}{c|}{Est SE} &
\multicolumn{3}{c|}{Emp SE}\\
\cline{3-11}
&  &$\beta_1$  &$\beta_2$ &$\beta_3$ &$\beta_1$  &$\beta_2$ &$\beta_3$&$\beta_1$  &$\beta_2$ &$\beta_3$\\
\hline
30\%&0,0,0.1	&0.01	&0.02	&-0.01	&0.20	&0.06	&0.03	&0.18	&0.10	 &0.05\\
	&0,0,0.2	&0.01	&0.02	&-0.01	&0.26	&0.06	&0.04	&0.20	&0.11	 &0.06\\
	&0,-0.5,0.1	&0.01	&0.01	&0.00	&0.24	&0.07	&0.03	&0.20	&0.10	 &0.04\\
	&0,-0.5,0.2	&0.01	&0.01	&0.00	&0.22	&0.07	&0.03	&0.21	&0.11	 &0.05\\
	&0.5,0,0.1	&0.01	&0.02	&-0.01	&0.18	&0.05	&0.03	&0.18	&0.10	 &0.06\\
	&0.5,0,0.2	&0.00	&0.02	&-0.01	&0.20	&0.06	&0.05	&0.19	&0.10	 &0.07\\
	&0.5,-0.5,0.1	&0.00	&0.01	&0.00	&0.22	&0.07	&0.03	&0.20	 &0.10	&0.05\\
	&0.5,-0.5,0.2	&0.00	&0.01	&-0.01	&0.22	&0.09	&0.05	&0.21	 &0.11	&0.06\\
\hline
50\%&0,0,0.1	&0.01	&0.02	&0.00	&0.18	&0.06	&0.03	&0.20	&0.12	 &0.07\\
	&0,0,0.2	&0.02	&0.01	&-0.01	&0.27	&0.08	&0.05	&0.22	&0.13	 &0.08\\
	&0,-0.5,0.1	&0.01	&0.01	&0.00	&0.22	&0.07	&0.03	&0.22	&0.13	 &0.06\\
	&0,-0.5,0.2	&0.01	&0.01	&0.00	&0.23	&0.07	&0.04	&0.24	&0.14	 &0.07\\
	&0.5,0,0.1	&0.01	&0.01	&0.00	&0.13	&0.05	&0.04	&0.19	&0.12	 &0.08\\
	&0.5,0,0.2	&0.00	&0.01	&0.00	&0.24	&0.07	&0.06	&0.20	&0.12	 &0.10\\
	&0.5,-0.5,0.1	&0.01	&0.01	&0.00	&0.16	&0.07	&0.03	&0.21	 &0.12	&0.07\\
	&0.5,-0.5,0.2	&0.02	&0.01	&0.00	&0.18	&0.07	&0.04	&0.22	 &0.13	&0.08\\
\hline
\end{tabular}
\end{center}
\tabcolsep 5mm \caption{Simulation results for sample size 500 for the baseline hazard $1/(1+t)$, binary covariates, with different true $\bmbeta$ and censoring rate from 100 simulations}
\end{table}

We found that the numerical variance estimator is not stable even for large sample size. This might be due to the fact that the numerical method is essentially smoothing with uniform kernel and the kernel estimator is not stable with finite sample size. But we found that the variance can be estimated precisely with Bootstrap technique even for a relatively small sample size (n=200). Thus in data analysis, though computational extensive, we recommend using Bootstrap for variance estimation.

\section{Real Data Examples}
\label{sec:4}
In this section, we illustrate our method by applying it to two data sets.
\subsection{Example 1: VA Lung Cancer}
A well known data set in survival analysis is the Veterna's Administration lung cancer trial data from Kalbfleisch and Prentice (1980). The covariates used are the Karnofsky score and the cell types. Because the additive effect magnitude depends on the time scale, here we choose 100 days as one unit in the analysis. Previous analysis showed that the proportional hazard assumption does not hold for Karnofsky score. According to our data, the proportional hazard assumption holds for some cell types. Thus we still model Karnofsky score with a multiplicative effect, but assumed cell types to have different kinds of effects. For illustration purpose, we used large and squamous types as reference group and fit an accelerated effect for the small type and fit an additive hazard effect for the adeno type. The fitted coefficients are given in Table 3 and the results suggest all three effects are significant.
\begin{table}[!h]
\begin{center}
\begin{tabular}{|c|c|c|c|c|c|}
\hline
Variable& Effect Type &Coef &SE &\multicolumn{2}{c|}{95\% CI}\\
\hline
Small type &Time Scale &0.94 &0.19 &0.55 &1.31\\
Karnofsky score (20) &Multiplicative &-0.75 &0.21 &-1.28 &-0.51\\
Adeno type &Additive &0.09 &0.04 &0.01 &0.18\\
\hline
\end{tabular}
\end{center}
\tabcolsep 5mm \caption{Result for VA Lung Cancer Data}
\end{table}
\subsection{Example 2: HIVNET 012}
We applied our method to the HIV-NET012 data. We dichotomized the maternal CD4+ counts and virus load at cut point 350 and 50,000. Here we use the time unit as 1,000 days. We first fit models that restrict all covariates to have the same type of effects, i.e. AFT model, Cox model or additive hazard model. Then we put in variables of our primary interest, which are the treatment in Cox part and virus load in time scale part and CD4+ counts in additive part. From Table 4, we see that the effect of Nevirapine is not statistically significant in all the models assuming same type of effects for virus load and CD4+ counts adjustment. But our proposed model shows that Nevirapine has a significant improvement in reducing hazard compared to AZT.
\begin{table}[!h]
\begin{center}
\begin{tabular}{|c|c|c|c|c|c|}
\hline
Variable& Effect Type &Coef &SE &\multicolumn{2}{c|}{95\% CI}\\
\hline
High VL &Time Scale &1.89 &0.42 &1.06 &2.72\\
Not use Nevirapine &Time Scale &0.35 &0.39 &-0.42 &1.12\\
Low CD4 &Time Scale &0.46 &0.37 &-0.27 &1.20\\
\hline
High VL &Multiplicative &0.98 &0.20 &0.58 &1.37\\
Not use Nevirapine &Multiplicative &0.18 &0.20 &-0.22 &0.57\\
Low CD4 &Multiplicative &0.24 &0.19 &-0.13 &0.62\\
\hline
High VL &Additive &0.14 &0.03 &0.08 &0.20\\
Not use Nevirapine &Additive &0.03 &0.03 &-0.03 &0.08\\
Low CD4 &Additive &0.03 &0.02 &-0.02 &0.07\\
\hline
High VL &Time Scale &2.05 &1.23 &-2.48 &2.71\\
Not use Nevirapine &Multiplicative &0.68 &0.32 &0.05 &1.34\\
Low CD4 &Additive &0.02 &0.02 &0.00 &0.06\\
\hline
\end{tabular}
\end{center}
\tabcolsep 5mm \caption{Result for HIV-NET Data}
\end{table}
The fitted curve is plotted in Figure 1 in comparison with the KM curve for each subgroup defined by virus load and CD4+ counts level. The time scales are different between the plots at left and those at right because we have a large estimated time scale change parameter. When looking at the Kaplan-Meier for longer period of time for the two groups with high virus load, there are signs of effect modification by time. However, when adding that interaction term into the model, its coefficient is -0.42 and is not statistically significant.

\begin{figure}
\caption{Comparing model fitted survival curve with the Kaplan-Meier Curve for each subgroup defined by virus load and CD4+ level: red line is model fitted; black line is Kaplan-Meier; solid line is nevirapine group; dotted line is AZT}
\centering
\includegraphics[width=130mm, trim=100 0 0 0]{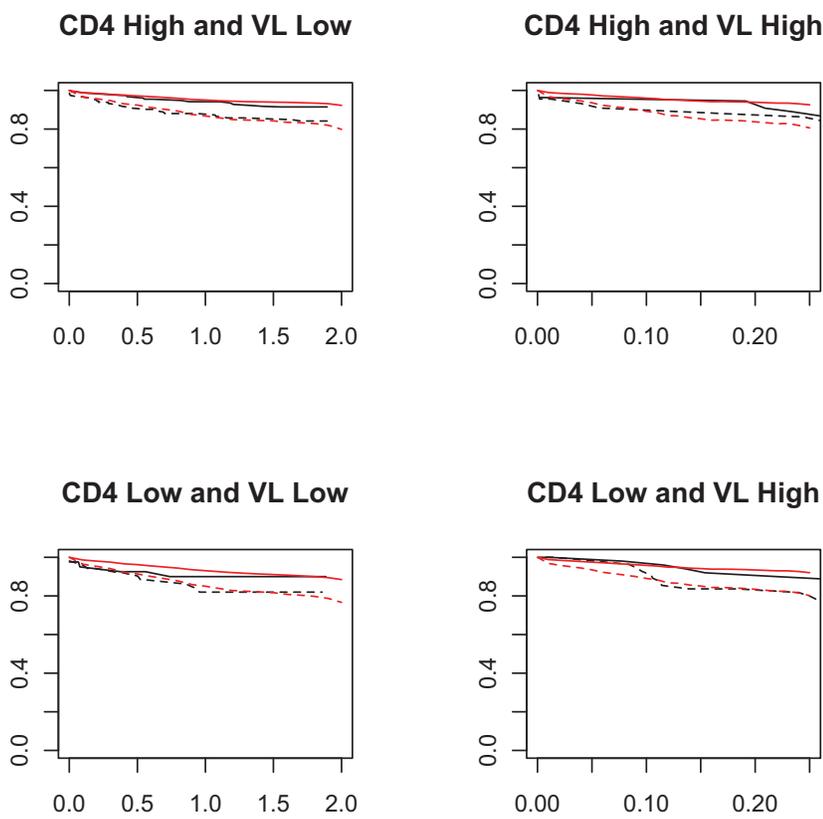}
\end{figure}

\section{Discussion}
\label{sec:5}
\subsection{Same covariate for different effects}
In this paper, we consider the case where $\bmZ_1$, $\bmZ_2$ and $\bmZ_3$ are different. In general, we can allow them to be the same. But we need to use different weight to estimate the effects. When the $\bmZ$s are the same for the three parts, the model is a higher level model that includes Cox model, AFT model and additive risk model. We can use this shape invariant model to test the goodness of fit for these models. When there are some common covariates in different parts of the model, the interpretation is slightly different. The model can have different signs for $\beta_2$, $\beta_3$ to allow cross over for the survival curve in different groups. In fact, even in real data, there are some examples where we expect the signs to be different. In this case, there are different pathways and the effects are in opposite sign in these pathways. However, there are some identifiability concerns when $\bmZ$s are the same. When the baseline hazard follows Weibull distribution with hazard $\lambda(x)=Cx^K$, if we have univariate $Z$, then the above quantities become $C(K+1)t^KZe^{[(K+1)\beta_1+\beta_2]Z}$, $Ct^KZe^{[(K+1)\beta_1+\beta_2]Z}$ and $Z$. Since the first two terms are exactly collinear, the model is not identifiable Also, the model is not identifiable when $Z$ is univariate and the true $\beta_1=0$, since the three quantities become $[\lambda_0(t)+t\dot{\lambda}_0(t)]Ze^{\beta_2Z}$, $\lambda_0(t)Ze^{\beta_2Z}$ and  $Z$ where the first two are collinear. To verify the identifiability for univariate $Z$ in general is difficult and simulation shows that there is multiple roots problem under moderate sample size (n=1000).

\subsection{General model form}
In fact, the model can be written in the following general forms:
\begin{equation}
\lambda(t\mid \bmZ)=\lambda_0(t l(\bmbeta_1 \bmZ_1))l(\bmbeta_1\bmZ_1)h(\bmbeta_2\bmZ_2(t))+g(\bmbeta_3\bmZ_3(t)),
\end{equation}
where $l(\cdot)$, $h(\cdot)$ and $g(\cdot)$ are known functions. The moment based estimator can be derived from this model exactly the same way as our special case discussed in the main text. But with different choice of $l(\cdot)$, $h(\cdot)$ and $g(\cdot)$, the interpretation of the parameters will be different and $\lambda_0(t)$ might no longer be baseline hazard. The estimating equation can be written as below:
\bea
&&\hat{\Lambda}_0(t;\bmbeta)\nonumber\\
&=&\int_0^t\frac{\sum_i[dN_i(s/l(\bmbeta_1\bmZ_{1i}))-Y_i(s/l(\bmbeta_1\bmZ_{1i}))g(\bmbeta_3\bmZ_{3i}(s))/l(\bmbeta_1\bmZ_{1i})ds]}{\sum_iY_i(s/l(\bmbeta_1\bmZ_{1i}))h(\bmbeta_2\bmZ_{2i}(s))}.
\eea
\bea
0&=&\sum_i\int\{[W_i(t)-\bar{W}(t;\bmbeta)]\nonumber \\
&&[dN_i(t/l(\bmbeta_1\bmZ_{1i}))-Y_i(t/l(\bmbeta_1\bmZ_{1i}))g(\bmbeta_3\bmZ_{3i}(t))/l(\bmbeta_1\bmZ_{1i})dt]\},
\eea
where $\bar{W}(t;\bmbeta)=\frac{\sum_jW_j(t)Y_j(t/l(\bmbeta_1\bmZ_{1i}))h(\bmbeta_2\bmZ_{2j}(t))}{\sum_jY_j(t/l(\bmbeta_1\bmZ_{1i}))h(\bmbeta_2\bmZ_{2j}(t))}$.

\subsection{Limitations}
One limitation for the moment based method is the potential loss of efficiency. For the moment estimating method, it is suggested that using the partial derivative of log hazard as weight function will gain more efficiency. However, such weight contains baseline hazard and its derivative; thus, it is not easy to estimate. Also numerical study shows that even we plugged in the true baseline to estimate weight, there still can be efficiency loss comparing to the simple weight $\bmZ$ since the estimating equation for baseline hazard is not the optimal one as derived from nonparametric maximum likelihood estimator (NPMLE). Merely changing weight for those equation for $\bmbeta$ does not ensure optimal efficiency. Besides efficiency issue, another limitation for the current method is the use of recursive method to find solution of estimating equations. The algorithm is slow when sample size is large and the method cannot be applied to high dimension covariates. One thing that affects the speed is that for this model the $W(t)$ and $\bar{W}(t)$ need to be calculated for each time $t_i\exp\{\beta_1Z_j\}$. And also, we notice that when the sample size is too small, there is severe convergence problem, which suggests that enough data is required to consider such a complex model.

Another limitation of our method is that we assume $\bmZ_1$ to be time independent in our general model (2). In real application, we might have $\bmZ_1$ be time-varying. However, there are some issues if we have a time-varying $\bmZ_1$: (1) The interpretation of $\bmbeta_1$ is difficult since the cumulative hazard will have a rather complicated form; (2) Estimation is not reliable since the proof for the asymptotic result cannot be generalized to the case where $\bmZ_1$ is time-varying.


\section{Technique Details}
Here we give the assumptions we need for the main theory as follows:
\begin{itemize}
\item We assume the integral in equation (9) is calculated under some truncation at time $\tau$ which satisfies $P(X_i>\tau e^{\bmbeta_{10}+\zeta})\geq \psi>0$ for all $i$. If we can find a $\tau$ such that conditions (A), (B) and (D) holds, then this assumption automatically hold.
\item (A) The baseline hazard function $\lambda_0(x)$ and the density function $f(x)=dF(x)/dx$ exists and are bounded by $K_1$ for some $\zeta>0$ for all $x<\tau e^{\bmbeta_{10}+\zeta}$.
\item (B) The density of censoring variable $C$, $h(x)=-dH(x)/dx$ exists and is bounded by $K_2$ for all $x<\tau e^{\bmbeta_{10}+\zeta}$.
(A) and (B) together imply that the density for variable $X$ exists and is bounded by $K=K_1+K_2$ for all $X<\tau e^{\bmbeta_{10}+\zeta}$.
\item (C) There are function $\theta(u)$ and a $\xi>0$ such that
\bea
|\lambda(u+\epsilon)-\lambda(u)-\epsilon\dot{\lambda}(u)|\leq \epsilon^2\theta(u) \nonumber
\eea
for $u\leq \tau e^{\bmbeta_{10}}$ and $\epsilon\leq \xi$ and 
\bea
\int_0^{\tau e^{\bmbeta_{10}}}|\theta(u)|du<\infty. \nonumber
\eea
\item (D) The absolute value of the covariate $\bmZ_i$ and the weight $W_i(t)$ are bounded by 1 for all subject $i$ on $[0,\tau e^{\bmbeta_{10}+\zeta}]$.
\item (E) There is a continuous function $\mu(u,\bmbeta)$ such that
\bea
\sup_{\bmbeta \in B(\bmbeta_0),u\leq \tau e^{\bmbeta_{10}+\zeta}}|\bar{W}(u,\bmbeta)-\mu(u,\bmbeta)|\longrightarrow_p 0. \nonumber
\eea
\item (F) Let
\bea
&&G(\bmbeta)\nonumber\\
&=&\lim_{n\longrightarrow \infty}\int_0^{\tau}n^{-1}\sum_{i=1}^n(W_i(u)-\bar{W}(u,\bmbeta))Y_i(ue^{-\bmbeta_{10} \bmZ_{1i}})\nonumber\\
&&\{-(ue^{(\bmbeta_{10}-\bmbeta_1)\bmZ_{1i}}\dot{\lambda}_0(ue^{(\bmbeta_{10}-\bmbeta_1)\bmZ_{1i}})+\lambda_0(ue^{(\bmbeta_{10}-\bmbeta_1)\bmZ_{1i}}))e^{\bmbeta_{20}\bmZ_{2i}(u)+(\bmbeta_{10}-\bmbeta_1)\bmZ_{1i}}\bmZ_{1i}\nonumber\\
&&-(\bmbeta_{30}-\bmbeta_{3})\bmZ_{3i}(u)e^{-\bmbeta_1\bmZ_{1i}}\bmZ_{1i}, -\lambda_0(u)\bmZ_{2i}(u)e
^{\bmbeta_{20}\bmZ_{2i}(u)}, -\bmZ_{3i}(u)e^{-\bmbeta_{10}\bmZ_{1i}}du\}.\nonumber
\eea
There is a matrix of continuous function $G(\bmbeta)$ such that
$$
\sup_{\bmbeta\in B(\bmbeta_0),u\leq \tau e^{\bmbeta_{10}+\zeta}}\Vert G(\bmbeta)-\int_0^{\tau}n^{-1}\sum_{i=1}^nG_i(u)du\Vert\longrightarrow_p 0.
$$
where the norm here represents the absolute value for each element.
\item (G) $W_i(t)$ converges uniformly in probability to a measurable functions $f(Z_i,t,\bmbeta,\lambda_0,data)$ respect to $\bmZ_i$, i.e.
\bea
\sup_{\bmbeta\in B(\bmbeta_0),t\in [0,\tau e^{\beta_{10}+\zeta}]}\Vert W_i-f(\bmZ_i,t,\bmbeta,\lambda_0,\underline{X})\Vert\longrightarrow_p 0.\nonumber
\eea
We denote $W_i^{\ast}(t)=f(Z_i,t,\bmbeta,\lambda_0)$ for simplicity. We assume $W_i^{\ast}(t)$ is a continuous function.
\end{itemize}

To develop the asymptotic results for our estimator, we follow the proof steps given in Tsiatis (1990). Denote
\bea
&&dM_i(u,\bmbeta) \nonumber\\
&=&dN_i(ue^{-\bmbeta_1\bmZ_{1i}})-Y_i(ue^{-\bmbeta_1\bmZ_{1i}})\lambda_0(ue^{(\bmbeta_{10}-\bmbeta_1)\bmZ_{1i}})e
^{\bmbeta_{20}\bmZ_{2i}(u)+(\bmbeta_{10}-\bmbeta_1)\bmZ_{1i}}du \nonumber\\
&&-Y_i(ue^{-\bmbeta_1\bmZ_{1i}})\bmbeta_{30}\bmZ_{3i}(u)e^{-\bmbeta_1\bmZ_{1i}}du.
\eea
We can write the score function $S_n(\bmbeta,\tau)$ as
\bea
&&\sum_{i=1}^n\int_0^{\tau} (W_i(u)-\bar{W}(u,\bmbeta))dM_i(u,\bmbeta) \nonumber\\
&&+\sum_{i=1}^n\int_0^{\tau}(W_i(u)-\bar{W}(u,\bmbeta))Y_i(ue^{-\bmbeta_1 \bmZ_{1i}}) \nonumber\\
&&\{\lambda_0(ue^{(\bmbeta_{10}-\bmbeta_1)\bmZ_{1i}})e
^{\bmbeta_{20}\bmZ_{2i}(u)+(\bmbeta_{10}-\bmbeta_1)\bmZ_{1i}}-\lambda_0(u)e
^{\bmbeta_2\bmZ_{2i}(u)}\nonumber \\
&&+(\bmbeta_{30}-\bmbeta_3)\bmZ_{3i}(u)e^{-\bmbeta_1\bmZ_{1i}}\}du\nonumber,
\eea
where the second term can be approximated by a linear form $nG(\bmbeta_{0})(\bmbeta-\bmbeta^0)$. Here $G(\cdot)$ is defined in Appendix assumption (F). So the score $S(\bmbeta)$ can be approximated by
\bea
\tilde{S}(\bmbeta)=S(\bmbeta_0)+nG(\bmbeta_0)(\bmbeta-\bmbeta_0)^T+o_p(1).
\eea
We define $\hat{\bmbeta}$ as the value that minimize the norm of the score $\Vert S(\bmbeta) \Vert$. Here we choose the $L^{\infty}$ norm for proof. We can also use the $L^2$ norm and since the two norms control each other, the convergence results will be the same. We define $\bmbeta^{\ast}$ as the solution of $\tilde{S}(\bmbeta)=0$,then we know that $\sqrt{n}(\bmbeta^{\ast}-\bmbeta_0)=\sqrt{n}G(\bmbeta_0)^{-1}S(\bmbeta_0)+o_p(1)$, which has a variance in a sandwich form
\bea
V=G(\bmbeta_0)^{-1}E\{S(\bmbeta_0)S(\bmbeta_0)^T\}G(\bmbeta_0)^{-T}.
\eea
We just need to show $\sqrt{n}(\hat{\bmbeta}-\bmbeta^{\ast})\rightarrow_p 0.$ in order to get the asymptotic results for $\hat{\bmbeta}$. Proven by Jureckova (1969, 1971), it would suffice to prove that
$$
\sup_{\Vert \bmbeta-\bmbeta_0 \Vert\leq n^{-1/2}C}n^{-1/2}\Vert S(\bmbeta)-\tilde{S}(\bmbeta) \Vert\rightarrow_p 0,
$$
for any $C>0$, which can be derived in three parts. First, we show that for any fixed $d$, we have
$n^{-1/2}(S(\bmbeta_0+n^{-1/2}d)-\tilde{S}(\bmbeta_0+n^{-1/2}d))\rightarrow_p 0.$ Then we show uniform convergence at a fixed finite number of points that form a mesh from $-C$ to $C$. If we have a sequence of $d_0$, $d_1$, $\cdots$, $d_m$, we need show $\max n^{-1/2}\Vert S(\bmbeta_0+n^{-1/2}d_i)-\tilde{S}(\bmbeta_0+n^{-1/2}d_i) \Vert\rightarrow_p 0.$ and $n^{-1/2}S(\bmbeta)$ (as a function of $\bmbeta$) do not fluctuate too greatly within any interval of the mesh. Mathematically, if we have the mesh with size $\delta>0$, then we just need to show that for any $\epsilon>0$, there exist $\delta>0$ such that
$$
\lim_{n\rightarrow \infty}P\{\sup_{dn^{-1/2}\leq\Vert \bmbeta^{\ast}-\bmbeta_0\Vert\leq (d+\delta)n^{-1/2}}n^{-1/2}\Vert S(\bmbeta^{\ast})-S(\bmbeta_0+dn^{-1/2})\Vert\geq \epsilon\}=0.
$$

The proof of point-wise convergence is decomposed to the following three lemmas.

\textbf{Lemma 1} Let $\bmbeta_n=(\beta_{1n},\beta_{2n},\beta_{3n})$ denote a sequence of constant vectors converging to $\bmbeta_0$. Then
$$
n^{-1/2}\sum_{i=1}^n\left[\int_0^{\tau} (W_i(u)-\bar{W}(u,\bmbeta_n))dM_i(u,\bmbeta_n)-\int_0^{\tau} (W_i(u)-\mu(u,\bmbeta_n))dM_i(u,\bmbeta_n)\right]\nonumber
$$
converges to 0 in probability, where $\mu(\cdot)$ is defined in condition (E).

\textbf{Lemma 2}
$$
n^{-1/2}\{\sum_{i=1}^n\int_0^{\tau} (W_i(u)-\bar{W}(u,\bmbeta_n))dM_i(u,\bmbeta_n)-S(0)\}\longrightarrow_p 0.
$$

\textbf{Lemma 3}
\bea
&&n^{-1}\sum_{i=1}^n\int_0^{\tau}(W_i(u)-\bar{W}(u,\bmbeta_n))Y_i(ue^{-\bmbeta_{1n} \bmZ_{1i}})\nonumber\\
&&\{\lambda_0(ue^{(\bmbeta_{10}-\bmbeta_{1n})\bmZ_{1i}})e
^{\bmbeta_{20}\bmZ_{2i}(u)+(\bmbeta_{10}-\bmbeta_{1n})\bmZ_{1i}}-\lambda_0(u)e
^{\bmbeta_{2n}\bmZ_{2i}(u)}\nonumber\\
&&+(\bmbeta_{30}-\bmbeta_{3n})\bmZ_{3i}(u)e^{-\bmbeta_{n1}\bmZ_{1i}}\}du\nonumber\\
&=&\bmbeta_n(G(0)+o_p(1)).\nonumber
\eea

\textbf{Proof of Lemma 1:} The expression in the Lemma 1 equals to $R(\tau)$, where
\beaa
R(u)=n^{-1/2}\sum_{i=1}^n\int_0^{u} (\mu(u,\bmbeta_n)-\bar{W}(u,\bmbeta_n))dM_i(u,\bmbeta_n)
\eeaa
is a martingale. So by Lenglart's inequality (1977) given by Andersen and Gill (1982), we have that for each components of the $R(u)=(R_1(u),R_2(u),R_3(u))^T$,
\beaa
P(|R_k(u)|>\epsilon)&\leq& \delta/\epsilon^2+P(n^{-1}\sum_{i=1}^n\int_0^u (\mu_k(u,\bmbeta_n)-\bar{W}_k(u,\bmbeta_n))^2Y_i(ue^{-\beta_{1n}Z_i})\\
&&[\lambda_0(ue^{(\bmbeta_{10}-\bmbeta_{1n})\bmZ_{1i}})e
^{\bmbeta_{20}\bmZ_{2i}(u)+(\bmbeta_{10}-\bmbeta_{1n})\bmZ_{1i}}+\bmbeta_{30}\bmZ_{3i}(u)e^{-\bmbeta_{1n}\bmZ_{1i}}]du>\delta).
\eeaa
By the definition of $\bmbeta_n$, we can find $N$ sufficiently large such that $\bmbeta_N\in B(\bmbeta_0)$ and $\Vert \bmbeta_N-\bmbeta_0\Vert<\zeta$. Also, by assumption (E), we can find $N(\epsilon,K_k)$ such that  for any $n>N(\epsilon,K_k)$,
\beaa
P(\sup_{u\leq \tau}|\mu_k(u,\bmbeta_n)-\bar{W}_k(u,\bmbeta_n)|>K_k)<\epsilon /6.
\eeaa
So we have, with probability exceeding $1-\epsilon /6$,
\beaa
&&\int_0^\tau (\mu_k(u,\bmbeta_n)-\bar{W}_k(u,\bmbeta_n))^2Y_i(ue^{-\bmbeta_{1n}\bmZ_{1i}})\\
&&[\lambda_0(ue^{(\bmbeta_{10}-\bmbeta_{1n})\bmZ_{1i}})e
^{\bmbeta_{20}\bmZ_{2i}(u)+(\bmbeta_{10}-\bmbeta_{1n})\bmZ_{1i}}+\bmbeta_{30}\bmZ_{3i}(u)e^{-\bmbeta_{1n}\bmZ_{1i}}]du)\\
&&\leq K_k^2\Lambda_0(ue^{|\bmbeta_{10}-\bmbeta_{1n}|})e^{\bmbeta_{20}}+\bmbeta_{30}\tau.
\eeaa
For large $N$, we have $|\bmbeta_{10}-\bmbeta_{1N}|<\zeta$. By truncation assumption, $\Lambda_0(ue^{\bmbeta_{10}-\bmbeta_{1n}})=-\log (S_0(ue^{\bmbeta_{10}-\bmbeta_{1n}}))\leq -\log(\psi)$, so if we choose $K_k\leq \left[\frac{\delta}{-\log(\psi)+\bmbeta_{30}\tau}\right]^{1/2}$, we have
\beaa
&&P(n^{-1}\sum_{i=1}^n\int_0^u (\mu_k(u,\bmbeta_n)-\bar{W}_k(u,\bmbeta_n))^2Y_i(ue^{-\bmbeta_{1n}\bmZ_{1i}})\\
&&[\lambda_0(ue^{(\bmbeta_{10}-\bmbeta_{1n})\bmZ_{1i}})e
^{\bmbeta_{20}\bmZ_{2i}(u)}+\bmbeta_{30}\bmZ_{3i}(u)]e^{-\bmbeta_{1n}\bmZ_{1i}}du)<\epsilon /6.
\eeaa
By choosing $\delta=\epsilon^3/6$, we have $P(|R_k(u)|>\epsilon)<\epsilon /3$.
So we have
\beaa
P(\Vert R(u)\Vert>\epsilon)\leq \sum_{k=1}^3 P(|R_k(u)|>\epsilon)=\epsilon
\eeaa

\textbf{Proof of Lemma 2}:
\beaa
&&n^{-1/2}\{\sum_{i=1}^n\int_0^{\tau} (W_i(u)-\bar{W}(u,\bmbeta_n))dM_i(u,\bmbeta_n)-S(0)\}\longrightarrow_p 0\\
&=&n^{-1/2}\{\sum_{i=1}^n\int_0^{\tau} (W_i(u)-\bar{W}(u,\bmbeta_n))dM_i(u,\bmbeta_n)-\sum_{i=1}^n\int_0^{\tau} (W_i(u)-\mu(u,\bmbeta_n))dM_i(u,\bmbeta_n)\}\\
&&+n^{-1/2}\{\sum_{i=1}^n\int_0^{\tau} (W_i(u)-\mu(u,\bmbeta_n))dM_i(u,\bmbeta_n)-\sum_{i=1}^n\int_0^{\tau} (W_i(u)-\mu(u,\bmbeta_0))dM_i(u,\bmbeta_0)\}\\
&&+n^{-1/2}\{\sum_{i=1}^n\int_0^{\tau} (W_i(u)-\mu(u,\bmbeta_0))dM_i(u,\bmbeta_0)-\sum_{i=1}^n\int_0^{\tau} (W_i(u)-\bar{W}(u,\bmbeta_0))dM_i(u,\bmbeta_0)\}
\eeaa
The first and third terms converge to 0 in probability by Lemma 1, so we just focus the second term which we can further decompose to three terms
\beaa
&&n^{-1/2}\{\sum_{i=1}^n\int_0^{\tau} (W_i(u)-\mu(u,\bmbeta_n))dM_i(u,\bmbeta_n)-\sum_{i=1}^n\int_0^{\tau} (W_i(u)-\mu(u,\bmbeta_0))dM_i(u,\bmbeta_0)\}\\
&=&n^{-1/2}\{\sum_{i=1}^n\int_0^{\tau} (W_i(u)-\mu(u,\bmbeta_n))dM_i(u,\bmbeta_n)-\sum_{i=1}^n\int_0^{\tau} (W_i(u)-\mu(u,\bmbeta_n))dM_i(u,\bmbeta_0)\}\\
&&+n^{-1/2}\{\sum_{i=1}^n\int_0^{\tau} (W_i(u)-\mu(u,\bmbeta_n))dM_i(u,\bmbeta_0)-\sum_{i=1}^n\int_0^{\tau} (W_i(u)-\mu(u,\bmbeta_0))dM_i(u,\bmbeta_0)\}\\
&=&III+IV.
\eeaa
For term IV, it equals
\beaa
n^{-1/2}\{\sum_{i=1}^n\int_0^{\tau} (\mu(u,\bmbeta_0)-\mu(u,\bmbeta_n))dM_i(u,\bmbeta_0)
\eeaa
and by assumption (D), its variance is bounded by
\beaa
&&n^{-1}\sum_{i=1}^n\int_0^{\tau}(\mu(u,\bmbeta_0)-\mu(u,\bmbeta_n))^2\lambda_i(u)P(X_i\geq u)du\\
&=&n^{-1}\sum_{i=1}^n\int_0^{\tau}(\mu(u,\bmbeta_0)-\mu(u,\bmbeta_n))^2\lambda_i(u)S_i(x)H_i(x)du\\
&\leq &n^{-1}\sum_{i=1}^n\int_0^{\tau}(\mu(u,\bmbeta_0)-\mu(u,\bmbeta_n))^2\lambda_i(u)S_i(x)du\leq 4
\eeaa
So by dominated convergence theorem and continuity of $\mu(\cdot)$, when $\bmbeta_n\longrightarrow_p \bmbeta$, we have the term IV converge to 0 in probability. For term III, as we have
\beaa
dM_i(ue^{(\beta_{1n}-\beta_{10})\bmZ_i},\bmbeta)=dM_i(u,\bmbeta_0),
\eeaa
So term III can be written as
\beaa
&&n^{-1/2}\{\sum_{i=1}^n\int_0^{\tau e^{(\bmbeta_{1n}-\bmbeta_{10})\bmZ_{1i}}} (W_i(ue^{(\bmbeta_{1n}-\bmbeta_{10})\bmZ_{1i}})\\
&&-\mu(ue^{(\bmbeta_{1n}-\bmbeta_{10})\bmZ_{1i}},\bmbeta_n))dM_i(u,\bmbeta_0)-\sum_{i=1}^n\int_0^{\tau} (W_i(u)-\mu(u,\bmbeta_n))dM_i(u,\bmbeta_0)\}
\eeaa
This can be further decomposed to three terms $V+VI+VII$.
\beaa
V&=&n^{-1/2}\sum_{i=1}^n\int_0^{\tau} (\mu(u,\bmbeta_n)-\mu(ue^{(\bmbeta_{1n}-\bmbeta_{10})\bmZ_{1i}},\bmbeta_n))dM_i(u,\bmbeta_0)\\
VI&=&n^{-1/2}\sum_{i=1}^n\int_0^{\tau} (W_i(ue^{(\bmbeta_{1n}-\bmbeta_{10})\bmZ_{1i}})-W_i(u))dM_i(u,\bmbeta_0)\\
VII&=&n^{-1/2}\sum_{i=1}^n\int_{\tau}^{\tau e^{(\bmbeta_{1n}-\bmbeta_{10})\bmZ_{1i}}} (W_i(ue^{(\bmbeta_{1n}-\bmbeta_{10})\bmZ_{1i}})-\mu(ue^{(\bmbeta_{1n}-\bmbeta_{10})\bmZ_{1i}},\bmbeta_n))dM_i(u,\bmbeta_0).
\eeaa
Similar to arguments for IV, term V has a bounded variance 4 and thus converges to 0 in probability when $\bmbeta_n$ converges to $\bmbeta_0$ by dominated convergence theorem and continuity of $\mu(\cdot)$.
For term VI, by assumption (G), it is asymptotically equivalent to
\beaa
n^{-1/2}\sum_{i=1}^n\int_0^{\tau} (W_i^{\ast}(ue^{(\bmbeta_{1n}-\bmbeta_{10})\bmZ_{1i}})-W_i^{\ast}(u))dM_i(u,\bmbeta_0),
\eeaa
Its variance is
\beaa
&&n^{-1}\sum_{i=1}^n\int_0^{\tau}(W_i^{\ast}(ue^{(\bmbeta_{1n}-\bmbeta_{10})\bmZ_{1i}})-W_i^{\ast}(u))^2\lambda_i(u)P(X_i\geq u)du\\
&\leq &n^{-1}\sum_{i=1}^n\int_0^{\tau}4\lambda_i(u)S_i(u)du\leq 4
\eeaa
So VI converges to 0 in probability when $\bmbeta_n$ converges to $\bmbeta_0$ by continuity of $W_i^{\ast}$ and dominated convergence theorem.

For term VII, as both $Z$ and $W$ are bounded, it is asymptotically equivalent to
\beaa
n^{-1/2}\sum_{i=1}^n\int_{\tau}^{\tau e^{(\bmbeta_{1n}-\bmbeta_{10})\bmZ_{1i}}} (W_i^{\ast}(ue^{(\bmbeta_{1n}-\bmbeta_{10})\bmZ_{1i}})-\mu(ue^{(\bmbeta_{1n}-\bmbeta_{10})\bmZ_{1i}},\bmbeta_n))dM_i(u,\bmbeta_0)
\eeaa
whose variance is
\beaa
&&n^{-1}\sum_{i=1}^n\int_{\tau}^{\tau e^{(\bmbeta_{1n}-\bmbeta_{10})\bmZ_{1i}}}(W_i^{\ast}(ue^{(\bmbeta_{1n}-\bmbeta_{10})\bmZ_{1i}})-\mu(ue^{(\bmbeta_{1n}-\bmbeta_{10})\bmZ_{1i}},\bmbeta_n))^2\lambda_i(u)P(X_i\geq u)du\\
&\leq &4n^{-1}\sum_{i=1}^n\int_{\tau}^{\tau e^{\bmbeta_{1n}-\bmbeta_{10}}}f_i(x).
\eeaa
Since we have
\beaa
f_i(t)&\leq &\lambda_i(t)\\
&\leq & \exp\{\bmbeta_3t\}+\lambda_0(te^{\bmbeta_1\bmZ_1})e^{\bmbeta_1+\bmbeta_2}\\
&\leq & \exp\{\bmbeta_3t\}+K_1e^{\bmbeta_1+\bmbeta_2},
\eeaa
the variance of VII is smaller than
\beaa
&&4n^{-1}\sum_{i=1}^n\tau (e^{\bmbeta_{1n}-\bmbeta_{10}}-1)[\exp\{\bmbeta_3\tau e^{\bmbeta_{1n}-\bmbeta_{10}}\}+K_1e^{\bmbeta_1+\bmbeta_2}]\\
&\leq & C (\bmbeta_{n1}-\bmbeta_{10})\longrightarrow 0.
\eeaa
So by the weak law of large number, VII converges to 0 with $\bmbeta_n$ approaching $\bmbeta_0$. Combining the results above, we proved Lemma 2.

\textbf{Proof of Lemma 3}: We can decompose the quantity in Lemma 3 to the following parts.

\beaa
&&A_1=n^{-1}\sum_{i=1}^n\int_0^{\tau}(W_i(u)-\bar{W}(u,\bmbeta_n))Y_i(ue^{-\bmbeta_{1n} \bmZ_{1i}})\\
&&\{\lambda_0(ue^{(\bmbeta_{10}-\bmbeta_{1n})\bmZ_{1i}})e
^{\bmbeta_{20}\bmZ_{2i}(u)+(\bmbeta_{10}-\bmbeta_{1n})\bmZ_{1i}}-\lambda_0(u)e
^{\bmbeta_{20}\bmZ_{2i}(u)+(\bmbeta_{10}-\bmbeta_{1n})\bmZ_{1i}}\\
&&-u(e^{(\bmbeta_{10}-\bmbeta_{1n})\bmZ_{1i}}-1)\dot{\lambda}_0(u)e
^{\bmbeta_{20}\bmZ_{2i}(u)+(\bmbeta_{10}-\bmbeta_{1n})\bmZ_{1i}}\}du\\
&\leq &e^{\bmbeta_{20}+\bmbeta_{10}-\bmbeta_{1n}}(u(e^{(\bmbeta_{10}-\bmbeta_{1n})\bmZ_{1i}}-1))^2n^{-1}\sum_{i=1}^n\int_0^{\tau}(W_i(u)-\bar{W}(u,\bmbeta_n))Y_i(ue^{-\bmbeta_{1n} \bmZ_{1i}})\theta(u)du\\
&\leq &e^{\bmbeta_{20}+\bmbeta_{10}-\bmbeta_{1n}}(u(e^{(\bmbeta_{10}-\bmbeta_{1n})\bmZ_{1i}}-1))^2n^{-1}\sum_{i=1}^n\int_0^{\tau}(W_i(u)-\bar{W}(u,\bmbeta_n))^2\theta(u)du
\eeaa
Since $\theta(u)$ is integratable and $W_i$, $\bar{W}$ is bounded, $A_1\rightarrow_p 0$.
\beaa
&&A_2=n^{-1}\sum_{i=1}^n\int_0^{\tau}(W_i(u)-\bar{W}(u,\bmbeta_n))Y_i(ue^{-\bmbeta_{1n} \bmZ_{1i}})\\
&&\{u(e^{(\bmbeta_{10}-\bmbeta_{1n})\bmZ_{1i}}-1-(\bmbeta_{10}-\bmbeta_{1n})\bmZ_{1i}e^{(\bmbeta_{10}-\bmbeta_{1n})\bmZ_{1i}})\dot{\lambda}_0(u)e
^{\bmbeta_{20}\bmZ_{2i}(u)+(\bmbeta_{10}-\bmbeta_{1n})\bmZ_{1i}}\}du\\
&\leq &(\bmbeta_{10}-\bmbeta_{1n})^2n^{-1}\sum_{i=1}^n\int_0^{\tau}|W_i(u)-\bar{W}(u,\bmbeta_n)|Y_i(ue^{-\bmbeta_{1n} \bmZ_{1i}})|\dot{\lambda}_0(u)|e
^{\bmbeta_{20}+(\bmbeta_{10}-\bmbeta_{1n})}du
\eeaa
Since each term in the integral is bounded with probability 1, we have $A_2\rightarrow_p 0$.

\beaa
&&A_3=n^{-1}\sum_{i=1}^n\int_0^{\tau}(W_i(u)-\bar{W}(u,\bmbeta))Y_i(ue^{-\bmbeta_{1n} \bmZ_{1i}})\\
&&\{\lambda_0(u)e
^{\bmbeta_{2n}\bmZ_{2i}(u)}-\lambda_0(u)e
^{\bmbeta_{20}\bmZ_{2i}(u)}-\lambda_0(u)e
^{\bmbeta_{20}\bmZ_{2i}(u)}(\bmbeta_{2n}-\bmbeta_{20})\bmZ_{2i}\}du\\
&\leq &(\bmbeta_{2n}-\bmbeta_{20})^2n^{-1}\sum_{i=1}^n\int_0^{\tau}|W_i(u)-\bar{W}(u,\bmbeta)|\lambda_0(u)e
^{\bmbeta_{20}}du
\eeaa
and
\beaa
&&A_4=n^{-1}\sum_{i=1}^n\int_0^{\tau}(W_i(u)-\bar{W}(u,\bmbeta))Y_i(ue^{-\bmbeta_{1n} \bmZ_{1i}})\{(\bmbeta_{30}-\bmbeta_{3n})\bmZ_{3i}(u)e^{-\bmbeta_1\bmZ_{1i}}\}du
\eeaa
Since $W_i$ is bounded, it is easy to obtain both $A_3$ and $A_4$ converge to 0 in probability, so the Lemma 3 holds.

Combining Lemma 1-3 and let $\bmbeta_n=\bmbeta_0+n^{-1/2}d$, we proved the point-wise convergence. Then we count the number of sign interchanges caused by a small change in $\bmbeta$ to show that the score function converges uniformly. The score has partial derivative respect to $\bmbeta_2$ and $\bmbeta_3$, so the slight change in those two dimension will not cause problem. We just need to show that the change in $\bmbeta_1$ with $\delta$ size cause only bounded interchange and for each interchange, the change in statistics are bounded for any fixed $\bmbeta_2$ and $\bmbeta_3$ near the true $\bmbeta$. We rank the $\bmbeta_1\bmZ_{1i}$ and use same subscript $(i)$ for all variables.

The score can be written as
\beaa
S(\bmbeta)&=&\sum_{i}\int\{(W_{(i)}-\bar{W}_{(i)})[dN_{(i)}(te^{-\bmbeta_1 \bmZ_{1(i)}})-Y_{(i)}(te^{-\bmbeta_1 \bmZ_{1(i)}})\bmbeta_3\bmZ_{3(i)}(t)e^{-\bmbeta_1\bmZ_{1(i)}}dt]\}\\
&=&\sum_{i}(W_{(i)}-\bar{W}_{(i)})\Delta_{(i)}-\sum_{(i)}\int\{(W_{(i)}-\bar{W}_{(i)})[Y_{(i)}(te^{-\bmbeta_1 \bmZ_{1(i)}})\bmbeta_3\bmZ_{3i}(t) e^{-\bmbeta_1\bmZ_{1(i)}}dt]\},
\eeaa
where $\bar{W}_{(i)}(\bmbeta)=\bar{W}_{(i)}(\bmbeta, t_{(i)})$. The change of $\bmbeta$ causes change of $\bmbeta$ in two ways. We consider change $\bmbeta_1$ in the term $dN(\cdot)$ and $Y(\cdot)$ separately from that change in term like $\bmbeta_3\bmZ_{3i}(t)e^{\bmbeta_1Z_{1(i)}}$. As the score is derivable with respect to the parameter in the term like $\bmbeta_3\bmZ_{3i}(t)e^{\bmbeta_1Z_{1(i)}}$, we just need to show it is also bounded when it causes change in $dN(\cdot)$ and $Y(\cdot)$ but fixed the parameter in other parts.
We have $\bar{W_{(i)}}=\frac{\sum_{j>i}W_{(j)}(t_{(i)})e^{\bmbeta_2\bmZ_{2(j)}}}{\sum_{j>i}e^{\bmbeta_2\bmZ_{2(j)}}}$. So an interchange can cause the following quantity change in $\sum_{i}(W_{(i)}-\bar{W}_{(i)})\Delta_{(i)}$ similar to that derived by Tsiatis (3.15).
\beaa
&&\Delta_{(j+1)}\{W_{(j+1)}-\bar{W}_{(j)}\}+\Delta_{(j)}\{W_{(j)}-\frac{\bar{W}_{(j+2)}\sum_{i>j}e^{\bmbeta_2\bmZ_{2(i)}}+\bar{W}_{(j)}e^{\bmbeta_2\bmZ_{2(j)}}}{\sum_{i>j}e^{\bmbeta_2\bmZ_{2(i)}}}\}\\
&&-\Delta_{(j+1)}\{W_{(j+1)}-\bar{W}_{(j+1)}\}-\Delta_{(j)}\{W_{(j)}-\bar{W}_{(j)}\}\\
&=&\Delta_{(j+1)}(\bar{W}_{(j+1)}-W_{(j+1)})\frac{e^{\bmbeta_2\bmZ_{2(j+1)}}}{\sum_{i>j}e^{\bmbeta_2\bmZ_{2(i)}}}+\Delta_{(j)}(\bar{W}_{(j)}-\bar{W}_{(j+2)})\frac{\sum_{i>j+1}e^{\bmbeta_2\bmZ_{2(i)}}}{\sum_{i>j}e^{\bmbeta_2\bmZ_{2(i)}}}\\
&=&\{\Delta_{(j+1)}-\Delta_{(j)}\}(\bar{W}_{(j+1)}-W_{(j+1)}-\bar{W}_{(j)}+\bar{W}_{(j+2)})\frac{e^{\bmbeta_2\bmZ_{2(j+1)}-\bmbeta_1\bmZ_{1(j+1)}}}{\sum_{i>j}e^{\bmbeta_2\bmZ_{2(i)}-\bmbeta_1\bmZ_{1(i)}}}\\
&&+\Delta_{(j)}\frac{W_{(j+2)}e^{\bmbeta_2\bmZ_{2(j+2)}}+W_{(j+1)}e^{\bmbeta_2\bmZ_{2(j+1)}}}{\sum_{i>j}e^{\bmbeta_2\bmZ_{2(i)}}}.
\eeaa
By assumption (D), we have that $(\bar{W}_{(j+1)}-W_{(j+1)}-\bar{W}_{(j)}+\bar{W}_{(j+2)})$ is bounded by 4, $\Delta_{(j+1)}-\Delta_{(j)}$ is bounded by 2, and $\frac{e^{\bmbeta_2\bmZ_{2(j+1)}}}{\sum_{i>j}e^{\bmbeta_2\bmZ_{2(i)}}}$ bounded by $2e^{\bmbeta_2}/(N-j)$. So the overall change for the first part is bounded by $16e^{\bmbeta_2}/(N-j)+4e^{\bmbeta_2}/(N-j)=20e^{\bmbeta_2}/(N-j)$. By truncation assumption, $N-j$ should be in the same order as $N$ because more than $N\psi$ subject tend to be alive at $\tau e^{\bmbeta_{10}}$. So the quantity changed in score by one interchange is in the order of $O(1/N)$.

Now we calculate the change in second term
$$
\sum_{(i)}\int\{(W_{(i)}(u)-\bar{W}(u))[Y_{(i)}(e^{-\bmbeta_1 \bmZ_{1(i)}})\bmbeta_3\bmZ_{3(i)}(u)e^{-\bmbeta_1\bmZ_{1(i)}}du.
$$
As an integration of $Y(u)$, the change in $Y(u)$ only occurs at finite points, so
\beaa
\sum_{(i)}\int\{W_{(i)}(u)[Y_{(i)}(e^{-\bmbeta_1 \bmZ_{1(i)}})\bmbeta_3\bmZ_{3(i)}(u)e^{-\bmbeta_1\bmZ_{1(i)}}du
\eeaa
does not change. For the second term, it equals to
\beaa
-\int\{\bar{W}(u)\sum_i[Y_{(i)}(e^{-\bmbeta_1 \bmZ_{1(i)}})\bmbeta_3\bmZ_{3(i)}(u)e^{-\bmbeta_1\bmZ_{1(i)}}du.
\eeaa
For an interchange, the change in $\bar{W}(u)$ is uniformly bounded by $O(1/N)$ as shown in the first part. 

Same as Tsiatis (1990), we have
\beaa
\lim_{N\longrightarrow \infty}P(N^{-3/2}M\leq \epsilon)=0,
\eeaa
if we choose $\delta=O(\epsilon)$. Combining these results, we show the uniform convergence and thus obtain the asymptotic result we want.



\end{document}